\documentclass[12pt,preprint]{aastex}

\usepackage[lowtilde]{url}
\usepackage{amsmath}
\usepackage{color}

\newcommand {\Lya}    {Ly$\alpha$}   %  Lyalpha
\newcommand {\HI}        {\ion{H}{1}}   %  HI
\newcommand {\HeI}     {\ion{He}{1}}   %  HeI
\newcommand {\OIV}    {\ion{O}{4}}   %  OIV
\newcommand {\OV}    {\ion{O}{5}}    %  OV
\newcommand {\OVI}    {\ion{O}{6}}   %  OVI
\newcommand {\OVII}   {\ion{O}{7}}
\newcommand {\OVIII}  {\ion{O}{8}}

\newcommand {\CIV}    {\ion{C}{4}}
\newcommand {\CV}     {\ion{C}{5}}
\newcommand {\CVI}    {\ion{C}{6}}

\newcommand {\NV}     {\ion{N}{5}}
\newcommand {\NVI}     {\ion{N}{6}}
\newcommand {\NVII}    {\ion{N}{7}}

\newcommand {\NeVIII}  {\ion{Ne}{8}}   
\newcommand {\NeIX}  {\ion{Ne}{9}}

\newcommand {\cd}     {cm$^{-2}$}  
 
\newcommand {\HST}    {{\it HST}}

\newcommand {\etal}   {et~al.}

\begin{document}

\title{Identifying the Baryons in a Multiphase Intergalactic Medium}  

\author{J. Michael Shull \& Charles W. Danforth }
\affil{CASA, Department of Astrophysical \& Planetary Sciences, \\
University of Colorado, Boulder, CO 80309;  (303) 492-7827 }
\email{michael.shull@colorado.edu,   charles.danforth@colorado.edu}  

\noindent
{\bf In this white paper, we summarize the current observations of the baryon census at low redshift
(Shull, Smith, \& Danforth 2012).  
We then suggest improvements in measuring the baryons in major components  of the IGM and CGM 
with future UV and X-ray spectroscopic missions that could find and map the missing baryons,
the fuel for the formation and chemical evolution of galaxies.  }

\medskip

For low-redshift cosmology and galaxy formation rates, it is important to account for all the baryons synthesized 
in the Big Bang.   Cosmologists have noted a baryon deficit  in the low-redshift universe (Fukugita, Hogan, \& Peebles 
1998) relative to the predicted density synthesized in the Big Bang.  Although this deficit could arise from an incomplete 
inventory, it could also challenge our understanding of the thermodynamics of structure formation and the response of
the gas to accretion shocks and galactic outflows.  Recent analysis (Komatsu \etal\ 2011) of the spectrum of acoustic 
peaks in the Cosmic Microwave Background (CMB) obtained by the {\it Wilkinson Microwave Anisotropy Probe} (WMAP) 
found that baryons comprise a fraction $\Omega_b = 0.0455 \pm 0.0028$ of the critical matter-energy density of the
universe, $\rho_{\rm cr} = (9.205 \times 10^{-30}~{\rm g~cm}^{-3}) h_{70}^2$, where $h_{70}$ is the Hubble constant  
($H_0$) in units of 70 km~s$^{-1}$~Mpc$^{-1}$.  The product $\Omega_b \rho_{\rm cr}$ corresponds to a comoving  
hydrogen number density (at redshift $z =0$) of only $n_H = (1.9\times10^{-7} ~{\rm cm}^{-3}) h_{70}^2$.  

An inefficient distribution of collapsed baryons vs.\ distributed matter is a prediction of nearly all cosmological simulations 
(see Figure 1a) of large-scale structure formation (Cen \& Ostriker 1999, 2006;  Dav\'e \etal\ 1999, 2001; Smith \etal\ 2011; 
Tepper-Garcia \etal\ 2011).   These N-body hydrodynamical simulations suggest that 10--20\% of the baryons reside in 
collapsed objects and dense filaments, with the remaining 80\%  distributed over a wide range of phases in baryon overdensity 
($\Delta_b = \rho_b / \overline{\rho}_b)$ and temperature ($T$).   In fact, a shock-heated WHIM at $z < 1$ is a natural 
consequence of gravitational instability in a dark-matter dominated universe.  This hot gas is augmented by galactic-wind 
shocks and virialization in galaxy halos.  Together, these processes affect the rate of accretion onto galaxies
(cold-mode or  hot-mode) and control the process of galaxy and star formation.

%%%%%%%%%%%   Figure 1   %%%%%%%%%%%%%%%%%%%%%%%%%%%%%%%%

\begin{figure}
\epsscale{1.2} 
\plottwo {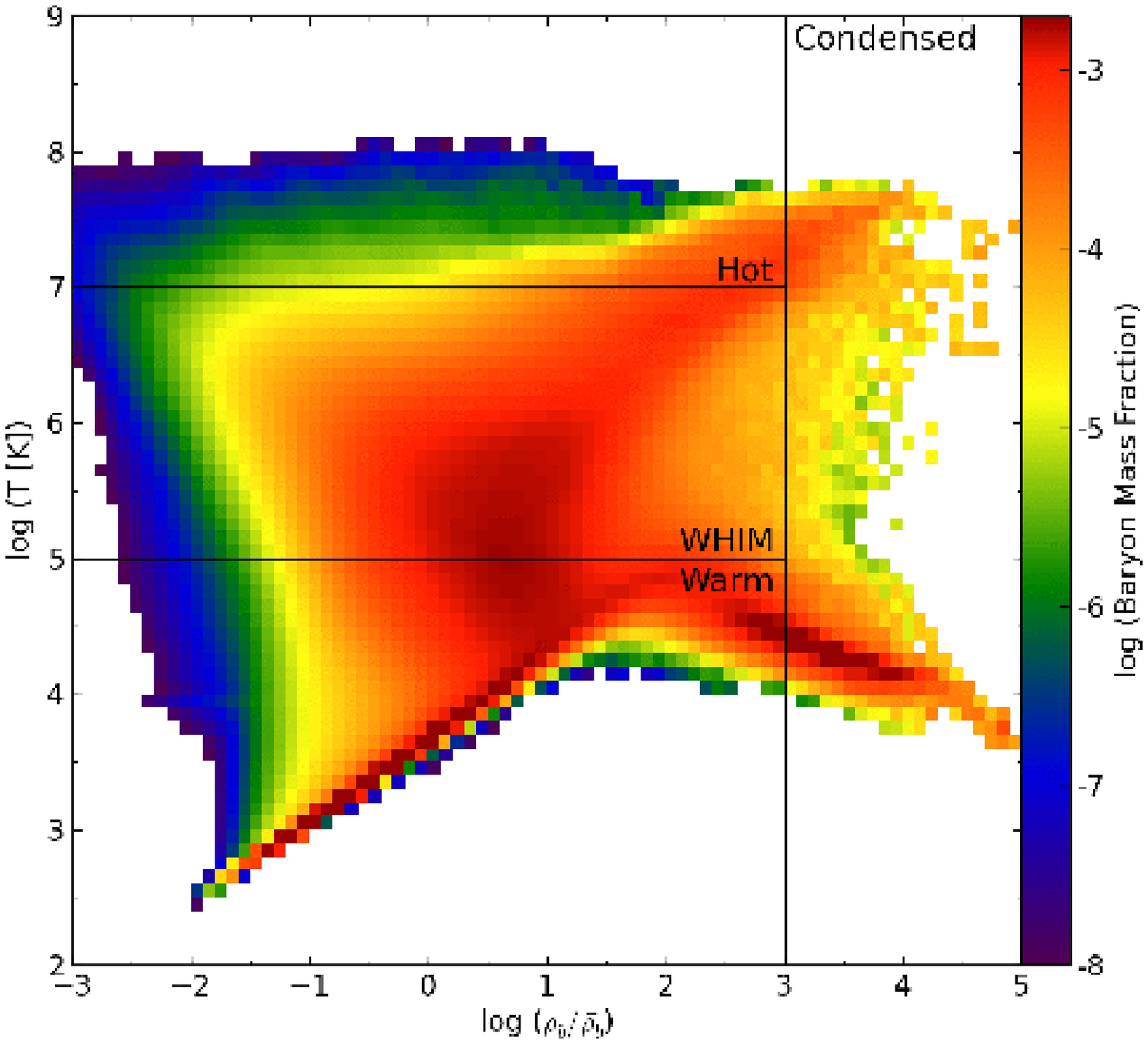} {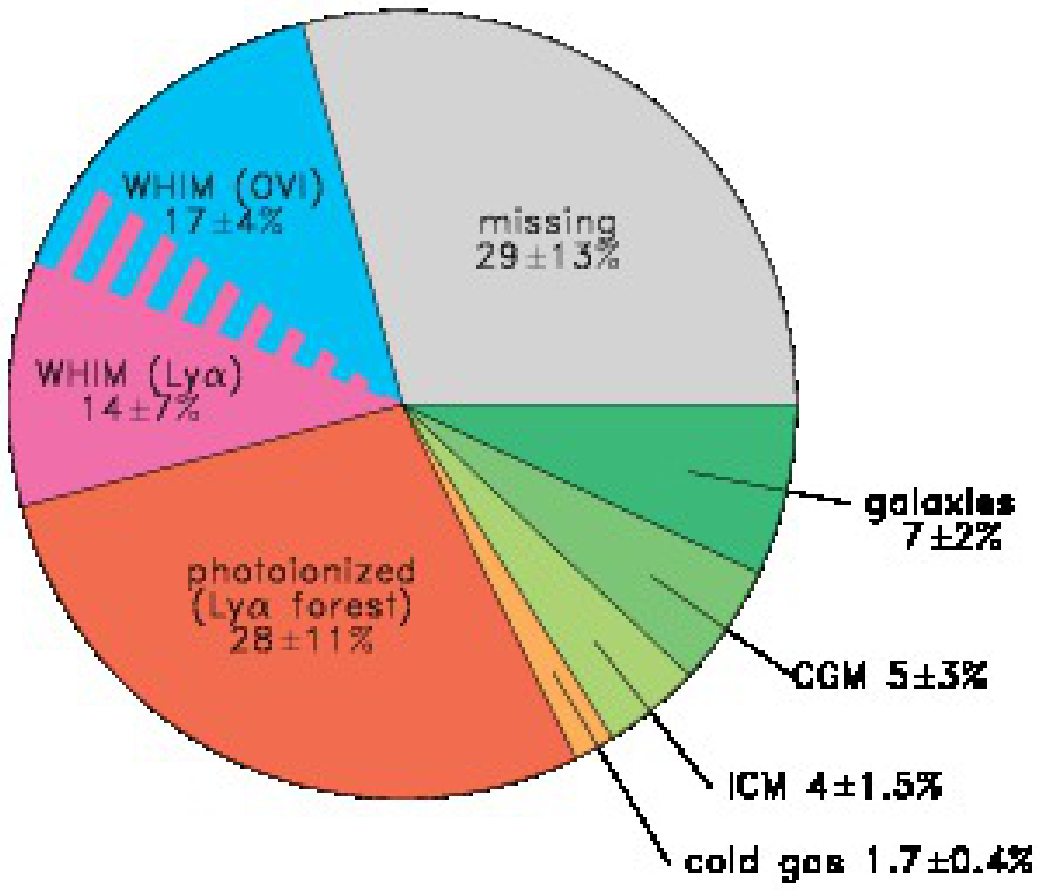} 
%   \plottwo {Fig1.eps} {Fig10.eps} 
\caption{Figures from recent baryon-census study by Shull, Smith, \& Danforth (2012). 
{\bf (Left)}  Simulated distribution of IGM in temperature and baryon overdensity 
   $\Delta_b = \rho_b / \overline{\rho}_b$, color-coded by baryon mass fraction.  
   Distribution shows the thermal phases commonly labeled as warm (diffuse photoionized gas), 
   WHIM (warm-hot intergalactic medium), and condensed.    
{\bf (Right)}  Compilation of current observational measurements of the low-redshift baryon census.
   Slices of pie-chart show baryons in collapsed form, in the circumgalactic medium (CGM) 
   and intercluster medium (ICM), and in cold gas (\HI\ and \HeI).  Primary baryon reservoirs include diffuse 
   photoionized \Lya\ forest and WHIM traced by \OVI\ and broad \Lya\ absorbers.  Collapsed phases
   (galaxies, CGM, ICM, cold neutral gas) total $18\pm4$\% and $29 \pm 13$\% of the baryons 
   remain unaccounted for.  An additional 15\%  may reside in X-ray absorbing gas at $T \geq 10^{6.3}$~K.   
   Additional baryons may be found in weaker lines of  low-column density \OVI\ and \Lya.
   Deeper spectroscopic UV and X-ray surveys are required to find and characterize this IGM and CGM,
   gas that provides fuel for new stars and galaxy formation.  
   }
\end{figure} 

%%%%%%%%%%%%%%%%%%%%%%%%%%%%%%%%%%%%%%%%%%%%%%%%

Unfortunately, the observed baryon inventories in the low redshift universe are uncertain.  Theoretical estimates of the
physical state of the gas are complicated by the formation of galaxies and large-scale structure and the  feedback from 
star formation in the form of ionizing radiation, metals, and outflows.  Galaxy surveys have identified $\sim$10\% of these 
baryons in collapsed objects such as galaxies, groups, and clusters (Salucci \& Persic 1999; Fukugita \& Peebles 2004).  
Over the last 15 years, substantial reservoirs of gas have also been found in the intergalactic medium (IGM), in the halos of 
galaxies, and in the circumgalactic medium (CGM) including metal-enriched gas blown out of galaxies
(Tumlinson \etal\ 2011; Prochaska \etal\ 2011).  Of the remaining 80-90\% of cosmological baryons, approximately half can 
be accounted for in the low-$z$ IGM (Bregman 2007;  Danforth \& Shull 2008) including the warm-hot IGM (or WHIM).   
Ultraviolet spectroscopic surveys of \Lya\ and \OVI\ have identified substantial numbers of absorbers 
(Danforth \& Shull 2008;  Tripp \etal\ 2008;  Thom \& Chen 2008), but claimed detections of hotter  in X-ray absorption by \OVII\  
(Nicastro \etal\ 2005a,b) remain controversial (Kaastra \etal\ 2006; Yao \etal\ 2012).  Unfortunately, X-ray spectra still have not 
confirmed the potential large reservoir of baryons at $T > 10^6$~K.

 Observations (Figure ab) of the ``Lyman-$\alpha$ forest" of absorption lines suggest that it contains $\sim$30\% of the 
 low-$z$ baryons (Penton \etal\ 2000, 2004; Lehner \etal\  2007; Danforth \& Shull 2008).  Another 30--40\% is 
 predicted by simulations  to reside in shock-heated gas at $10^5$~K to $10^7$~K (WHIM).   These two components 
 account for  60--70\% of the cosmological baryons.  However, owing to its low density, the WHIM is difficult to detect 
 in emission (Soltan 2006).  More promising are absorption-line studies that use the high ionization states of abundant 
 heavy elements with resonance lines in the far-ultraviolet (\CIV, \NV, \OVI),  extreme ultraviolet (\OIV, \OV, \NeVIII), 
and soft X-ray (\OVII, \OVIII, \NVI, \NeIX).   Gas in the temperature range $5 < \log T < 6$ can also be detected
in broad \Lya\ absorption (Richter \etal\ 2004; Danforth \etal\ 2010) arising from trace amounts of neutral hydrogen
with neutral fractions $-6.6 < \log f_{\rm HI} < -4.8$.  By far, the most effective surveys of the  low-$z$ WHIM were 
obtained from the \OVI\  lines at 1031.926~\AA\  and 1037.617~\AA\ (Danforth \& Shull 2008; Tripp \etal\ 2008; 
Thom \& Chen 2008), which probe the temperature range $10^{5.3-5.7}$~K in collisionally ionized gas.   Tilton \etal\
(2012) measured the column densities of 111 \OVI\  absorbers and estimated that $17\pm4$\% of the baryons reside in 
this phase, assuming new correction factors for the metallicity and \OVI\ ionization fraction (Shull \etal\ 2012).  
 A few detections of \NeVIII\ have also been reported (Savage \etal\ 2005, 2011;  Narayanan \etal\ 2009, 2011; 
 Meiring \etal\ 2012) probing somewhat hotter gas ($\log T \approx 5.7 \pm 0.2$).  

To detect even hotter portions of the WHIM at $\log T > 6$  requires X-ray searches for trace metal absorption lines 
from highly ionized C, O, or Ne.   Their weak X-ray absorption lines are difficult to detect with the 
current throughput and spectral resolution of spectrographs on {\it Chandra} and {\it XMM/Newton} (Yao \etal\ 2012).
Possible X-ray detections of hotter gas at $(1-3)\times10^6$~K have been claimed, using absorption lines of helium-like 
\OVII\  $\lambda21.602$ (Nicastro \etal\ 2005a,b, 2008;  Buote \etal\ 2009; Fang \etal\ 2010;  Zappacosta \etal\ 2010) and 
hydrogenic \OVIII\ $\lambda18.969$ (Fang \etal\ 2002, 2007).  Most of these {\it Chandra} detections remain 
controversial and unconfirmed by the {\it XMM-Newton} satellite (Kaastra \etal\ 2006; Williams \etal\ 2006; Rasmussen 
\etal\ 2007).  For example, recent analyses of spectroscopic data on Mrk~421 fail to detect any WHIM gas at the claimed redshifts 
($z = 0.01$ and 0.027), either in broad \Lya\ absorption (Danforth \etal\ 2011) from high-S/N data from the Cosmic Origins 
Spectrograph (COS) on the {\it Hubble Space Telescope} (\HST) or in \OVII\  (Yao \etal\ 2012) in {\it Chandra} data.

Figure 1b  shows a pie chart of the current observable distribution of low-redshift baryons in various 
forms, from collapsed structures to various phases of the IGM, CGM, and WHIM.   The slices show
contributions, $\Omega_b^{(i)} / \Omega_b^{(\rm tot)}$, to the total baryon content from
components ($i$).    Measurements of \Lya, \OVI, and broad \Lya\ absorbers,
together with more careful corrections for metallicity and ionization fraction, can now account for
$\sim60$\% of the baryons in the IGM.  An additional 5\% may reside in circumgalactic gas, 7\% 
in galaxies, and 4\% clusters.  {\it This still leaves a substantial fraction, $29\pm13$\%, unaccounted for.  } 
  
\noindent
{\bf  What observations and theoretical work are needed to make progress on the baryon census,
both in sensitivity and in accuracy? } First, we need more precise UV absorption-line surveys to measure
\OVI\ and \Lya\ absorbers to lower column densities.  The numbers of absorbers in current surveys become 
increasingly uncertain at column densities $N_{\rm HI} <  10^{13.0}$\cd\ and $N_{\rm OVI} < 10^{13.5}$\cd.
Finding and mapping this IGM/CGM fuel supply will require a new generation of spectrographs, optics,
and high-precision detectors  on a larger telescope ($D \geq 4^{\rm m}$ aperture;  $8^{\rm m}$ would be ideal).  
These weak-absorber surveys will require sensitivity to 2~m\AA\ equivalent widths, which is achievable 
at S/N = 50 toward many bright AGN background targets.     
We also need to obtain better detections and statistics for broad \Lya\ absorbers (BLAs) and 
the \NeVIII\ doublet (770.4, 780.3~\AA).  The \NeVIII\ lines are potentially more reliable probes of hot, 
collisionally ionized gas than \OVI, since \NeVIII\ requires 207~eV to produce and is likely to be less contaminated
by photoionization.   Redshifts $z > 0.47$ are needed to shift the \NeVIII\ lines into the \HST/COS band, but 
new far-UV missions with sensitivity down to 1000~\AA\ would open up many more AGN targets at $z > 0.30$.  
The BLAs also have considerable promise for WHIM probes, as they do not require corrections for metallicity.  
They do require determining the neutral fraction, $f_{\rm HI}$, through careful modeling of the gas temperature 
and ionization conditions.

It would also be helpful to verify the claimed X-ray detections of \OVII\ in the WHIM, most of which are not 
confirmed.   These new observations will allow us to explore the mixture of collisional ionization and 
photoionization in the WHIM, a project that requires understanding the implications of different
feedback mechanisms for injecting mass, thermal energy, and metals into the CGM.
How these metals mix and radiate likely determines the thermodynamics of the surrounding IGM.   
The most critical X-ray observations for the WHIM census will require a next generation of
spectrographs to measure the weak absorption lines of \OVII\ $\lambda21.602$, \OVIII\ $\lambda18.969$, 
and other He-like and H-like lines of abundant metals (\CV, \CVI, \NVI, \NVII).  As discussed by Yao \etal\ (2012), 
this requires high-throughput spectrographs ($E \approx 0.3-1.0$ keV) with energy resolution $E/\Delta E > 4000$ 
sufficient to resolve \OVII\  absorbers with m\AA\ equivalent width.  For weak lines, the predicted \OVII\
equivalent widths are $W_{\lambda} = (2.88~{\rm m\AA}) (N_{\rm OVII}/10^{15}~{\rm cm}^{-2})$.

%%%%%%%%%%%%%%%%%%%%%%%


\begin{references} 

{\small
\reference{Breg07}  Bregman, J. N. 2007, \araa, 45, 221

\reference {Buote09} Buote, D., Zappacosta, L., Fang, T.,  \etal\ 2009, \apj, 695, 1351

\reference {cenost99} Cen, R., \& Ostriker, J. P. 1999, \apj, 519, L109 

\reference {cenost06} Cen, R., \& Ostriker, J.  P. 2006, \apj, 650, 560
   
\reference{Dan08} Danforth, C. W., \& Shull, J. M. 2008, \apj, 679, 194 

\reference{Dan10} Danforth, C. W.,  Stocke, J. T., \& Shull, J. M.  2010,  \apj,  710, 613 

\reference{Dan11} Danforth, C. W.,  Stocke, J. T., Keeney, B. A., \etal\ 2011,  \apj, 743, 18

\reference{dave99}  Dav\'e, R.,  Hernquist, L., Katz, N., \& Weinberg, D. H.  1999, \apj,  511, 521

\reference{dave01}  Dav\'e, R., Cen, R., Ostriker, J. P.,  \etal\ 2001, \apj,  552, 473 

\reference{dave11}  Dav\'e, R.,  Oppenheimer, B. D., \& Finlator, K.  2011, \mnras, 415, 11

\reference {Fang02}  Fang, T.,  Marshall, H. L., Lee, J. C., David, D. S., \& Canizares, C. R. 2002, \apj, 572, L127

\reference {Fang07}  Fang, T., Canizares, C., \& Yao, Y.  2007, \apj, 670, 992

\reference {Fang10}  Fang, T., Buote, D. A., Humphrey, P. J.,  \etal\ 2010, \apj, 714, 1715 

\reference{Fuku98} Fukugita, M., Hogan, C. J., \& Peebles, P. J. E. 1998, \apj, 503, 518

\reference{Fuku04} Fukugita, M., \& Peebles, P. J. E. 2004, \apj, 616, 643 

\reference{Kaas06} Kaastra, J.,  Werner, N., den Herder, J. W.,  \etal\ 2006, \apj, 652, 189 

\reference{Kom11}  Komatsu, E.,  Smith, K. M., Dunkley, J., \etal\ 2011, \apjs, 192:18  

\reference{Lehn07} Lehner, N., Savage, B. D., Richter, P., \etal\  2007, \apj, 658, 680 

\reference{Meir12}  Meiring, J. D., Tripp, T. M., Werk, J. K., \etal\ 2012, \apj, submitted
   (arXiv:1201.0939)

\reference{Nar09} Narayanan, A., Savage, B. D., \& Wakker, B. P., \etal\ 2009, \apj, 703, 74

\reference{Nar11} Narayanan, A., Savage, B. D., Wakker, B. P., \etal\ 2011, \apj, 730, 15 

\reference{Nic05a}  Nicastro, F.,  Mathur, S., Elvis, M.,  \etal\ 2005a, \nat, 433, 495

\reference{Nic05b}  Nicastro, F., Mathur, S., Elvis, M.,  \etal\ 2005b, \apj, 629, 700

\reference{Pent00}  Penton, S. V., Shull, J. M., \& Stocke, J. T.  2000, \apj, 544, 150 

\reference{Pent04}  Penton, S. V., Stocke, J. T., \& Shull, J. M. 2004, \apjs, 152, 29

\reference{Pro11} Prochaska, J. X., Weiner, B., Chen, H.-W., Mulchaey, J., \& Cooksey, K. 2011,  
   \apj, 740, 91

\reference{Ras07} Rasmussen, A., Kahn, S. M., Paerels, F., \etal\ 2007, \apj, 656, 129

\reference{Richter04} Richter, P., Savage, B. D., Tripp, T. M., \& Sembach, K. R. 2004, 
   \apjs, 153, 165
   
\reference{Sal99} Salucci, P., \& Persic, M. 1999, \mnras, 309, 923

\reference{Sav05} Savage, B. D., Lehner, N., Wakker, B. P., Sembach, K., \& Tripp, T. 
   2005, \apj, 626, 776 
   
\reference{Sav11} Savage, B. D., Lehner, N., \& Narayanan, A.  2012 , \apj,  743, 180

\reference{Shu12} Shull, J. M., Smith, B. D.,  \& Danforth, C. W.   2012, \apj,  (arXiv:1112.2706)     

\reference{Smith10}  Smith, B. D.,  Hallman, E., Shull, J. M., \&  O'Shea, B. 2011, \apj, 731, 6

\reference{Solt06}  Soltan, A. M. 2006, A\&A, 460, 59 

\reference{Tepper11} Tepper-Garc{\'{\i}}a, T., Richter, P., Schaye, J., \etal\ 2011, \mnras, 413, 190 

\reference{ThomChen08} Thom, C., \& Chen, H.-W. 2008, \apj, 683, 22

\reference{Tilton12}  Tilton, E. M.,  Danforth, C. W., Shull, J. M., \& Ross, T. L. 2012, \apj, 
     (arXiv:1204.3623) 

\reference{Tripp08} Tripp, T.~M., Sembach, K.~R., Bowen, D.~V.,  \etal\ 2008, \apjs, 177, 39

\reference{Tumlin11} Tumlinson, J., Thom, C., Werk, J. K., \etal\ 2011, Science, 334, 998 

\reference{Will06} Williams, R., Mathur, S., Nicastro, F., \& Elvis, M.  2006, \apj, 642, L95

\reference {Yao12} Yao, Y., Shull, J. M., Wang, Q.-D., \& Cash, W. 2012, \apj, 746, 166

\reference {Zappa10} Zappacosta, L., Nicastro, F., Maiolino, R., \etal\ 2010, \apj, 714, 74 
}
\end{references}
\end{document}